\documentclass[twocolumn,aps,pra,superscriptaddress]{revtex4}
\usepackage{graphicx}
\usepackage{amsmath}
\usepackage{amssymb}
\usepackage{textcomp}
\usepackage{braket}
\usepackage[colorlinks=true,linkcolor=blue,citecolor=blue,urlcolor=blue]{hyperref}

\begin{document}

\title{Active Learning Algorithm for Computational Physics}

\author{Juan Yao}
\affiliation{Center for Quantum Computing, Peng Cheng Laboratory, Shenzhen, 518005, China}

\author{Yadong Wu}
\affiliation{Institute for Advanced Study, Tsinghua University, Beijing, 100084, China}

\author{Jahyun Koo}
\author{Binghai Yan}
\affiliation{Department of Condensed Matter Physics, Weizmann Institute of Science, Rehovot 76100, Israel }

\author{Hui Zhai}
\email{hzhai@tsinghua.edu.cn}
\affiliation{Institute for Advanced Study, Tsinghua University, Beijing, 100084, China}
\affiliation{Center for Quantum Computing, Peng Cheng Laboratory, Shenzhen, 518005, China}

\date{\today}
\begin{abstract}
In large-scale computation of physics problems, one often encounters the problem of determining a multi-dimensional function, which can be time-consuming when computing each point in this multi-dimensional space is already time-demanding. In the work, we propose that the active learning algorithm can speed up such calculations. The basic idea is to fit a multi-dimensional function by neural networks, and the key point is to make the query of labeled data economically by using a stratagem called ``query by committee". We present the general protocol of this fitting scheme, as well as the procedure of how to further compute physical observables with the fitted functions. We show that this method can work well with two examples, which are quantum three-body problem in atomic physics and the anomalous Hall conductivity in condensed matter physics, respectively. In these examples, we show that one reaches an accuracy of few percent error for computing physical observables with less than $10\%$ of total data points compared with uniform sampling. With these two examples, we also visualize that by using the active learning algorithm, the required data are added mostly in the regime where the function varies most rapidly, which explains the mechanism for the efficiency of the algorithm. We expect broad applications of our method on various kind of computational physics problems. 
\end{abstract}
\maketitle
\section{Background}
Neural network (NN) based supervised learning methods has nowadays found broad applications in studying quantum physics of condensed matter materials and atomic, molecular and optical systems \cite{review1,review2}. On the theoretical side, applications include, for example, finding orders and topological invariants in quantum phases \cite{Phase1,Phase2, Phase3, Phase4,Phase5,Phase6,Phase7}, generating variational wave functions for quantum many-body states \cite{VariationalWF, WFRBM, WF, WFFewBody, TheoryTomography} and speeding up quantum Monte Carlo sampling \cite{MC1, MC2}. On the experimental side, these methods can help both optimizing experimental protocols \cite{State, Tomography} and analyzing experimental data \cite{STM, ExpPhase, ExpFHM}. Usually the supervised learning scheme requires a huge set of labelled data. However, in many physics applications, labelling data can be quite expensive, for instance, performing computation or experiments repeatedly can be time- and resources-demanding. Therefore, in many cases labelled data are not abundant, which is a challenge that have prevented many applications. 

The active learning is a scheme to solve this problem \cite{ALbook}. It starts from training a NN with a small initial dataset, and then actively queries the labelled data based on the prediction of the NN and iteratively improves the performance of the NN until the goal of the task is reached. With this approach, sampling the large parameter space can be made more efficiently, and the request of labelled data is usually much more economical than normal supervised learning methods. Recently a few works have applied the active learning algorithm to determine the inter-atomic potentials in quantum materials \cite{PotentialAL1,PotentialAL2,PotentialAL3} and to optimize control in quantum experiments \cite{ExpControl1,ExpControl2}, where labelled data have to be obtained either by \textit{ab initio} calculation or by repeating experiments, which are both time consuming. 

In this work we focus on a class of general and common task in computational physics that is to numerically determining a multi-dimensional function, say, $\mathcal{F}(\alpha_1,\alpha_2,\dots,\alpha_n)$, where $\alpha_i$ are parameters.  Considering a uniform discretization, suppose we discretize each parameter into $L$ points, there are totally $L^{n}$ number of data points that need to be calculated. In many cases, calculation of each point already takes quite some time, and thus the total computational cost is massive. Nevertheless, for most functions, there are regimes where the function varies smoothly and regimes where the function varies rapidly. Ideally, one should sample more points in the steep regimes and less points in the smooth regimes in order to obtain a good fitting in the entire parameter space efficiently. However, it seems to be paradox because one does not know which regimes the function varies more rapidly prior to computing the function. 

Here we show that this goal can actually be achieved by using the active learning algorithm and the ``query by committee" stratagem \cite{Query_by_committee} to add data points iteratively. Below we will first introduce the general protocol and then demonstrate the algorithm in two concrete problems. One is the quantum three-body problem and the other is the anomalous Hall conductivity problem. These two are very representative examples in atomic and condensed matter physics, respectively.  Using these two examples, we will illustrate how the active learning and ``query by committee" guide adding data to steep regimes of the fitting function. We will also discuss how to compute physical observables from the fitted functions.

\begin{figure}[t]
\begin{center}
\includegraphics[width=0.45\textwidth]{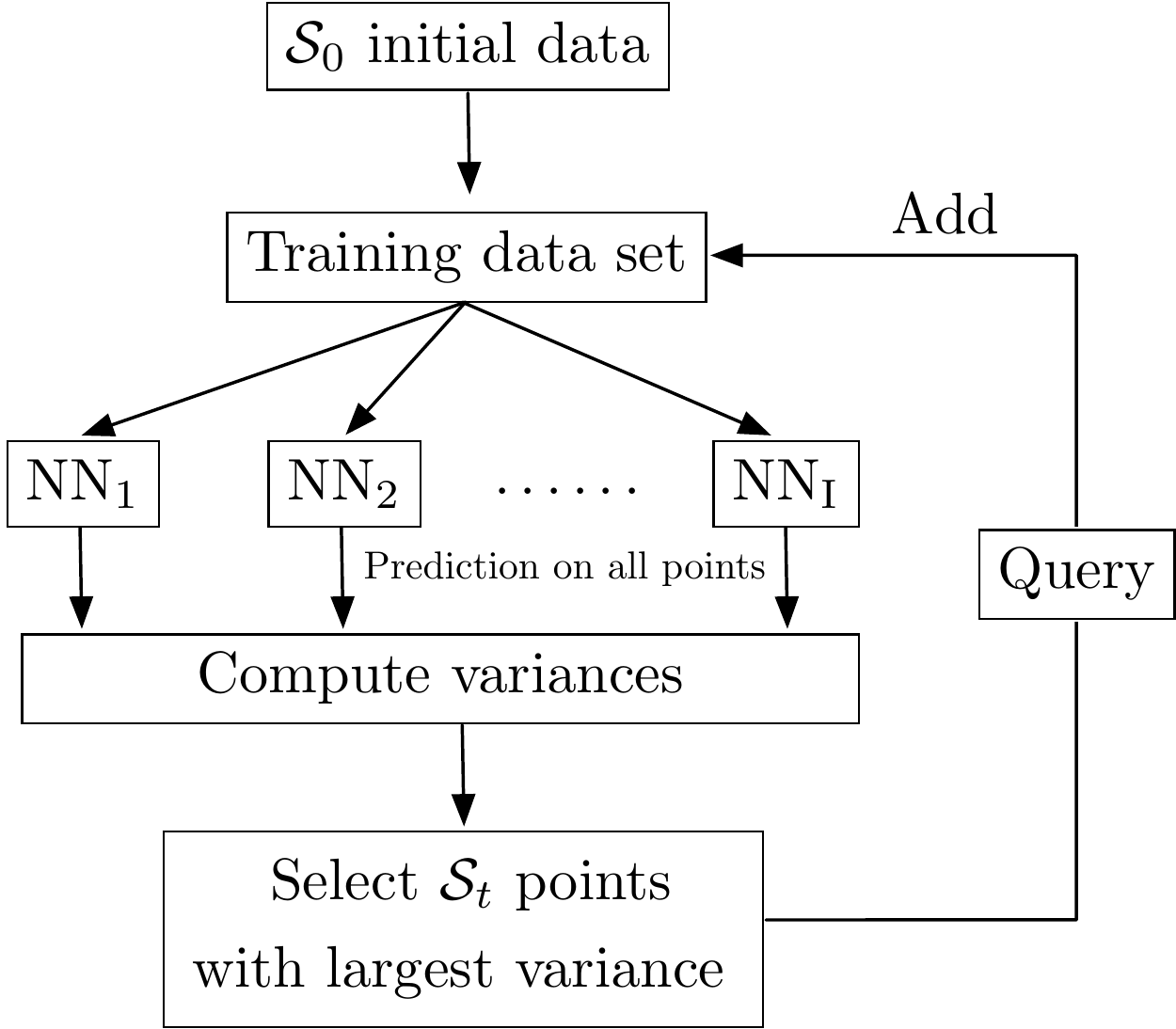}
\end{center}
\caption{Active learning protocol for fitting a multiply dimensional function. Here ``NN" represents ``neural network".  }
\label{protocal}
\end{figure}

\section{General Protocol.} 
The main procedure is summarized in Fig. \ref{protocal} and explained as follows:
 
1) We start with an initial dataset with the number of data $\mathcal{S}_0\ll L^n$, whose values of $\mathcal{F}$ have been computed exactly. We use this dataset to train a number of different NNs.

2) We ask all NNs to make predictions of $\mathcal{F}$ on all $L^n$ number of data points, and for each point we compute the variance among predications made by different NNs. 

3) We select $\mathcal{S}_t$ number of data point with the largest variance, again with $\mathcal{S}_t\ll L^n$, and we query the accurate value of $\mathcal{F}$ of these points by numerical calculation. 

4) We add the $\mathcal{S}_t$ number of new data into the training set to train all NNs again, and then repeat from step 2). We repeat the procedure for $m$-epochs until the results from all NNs converge. 

5) We use NN to calculate the value of $\mathcal{F}$ on all $L^n$ data points. 

In this protocol, one only needs to query the value of $\mathcal{F}$ on $\mathcal{S}_0+m\mathcal{S}_t$ number of data points, and we should keep $\mathcal{S}_0+m\mathcal{S}_t \ll L^n$. The trade-off is that we need to train a number of different NNs and keep using all NN to make predications on all data points. It can save computational cost if the computational cost for training NN and making predication with NN is much less than the computational cost of $\mathcal{F}$, which is often the case in many applications. 

In this protocol, the key idea is ``query by committee" \cite{Query_by_committee}. Here the committee is made of different NNs, which contain different number of layers, different number of nodes at each layer and different activation functions. Thanks to the great expressive power of NN, we do not need to assume a specific form of the fitting functions. In the regime where the function varies smoothly, different NNs can quickly reach a consensus and the variance will be small. On the other hand, in the regime where the function varies rapidly, it is hard for different NNs to converge and the variance will be large. Hence, we can use this variance to guide sampling data point. In fact, as we will see in the examples below, the data points added in later epoch are all added in the regime where the function changes rapidly.

\begin{figure}[t]
\begin{center}
\includegraphics[width=0.35\textwidth]{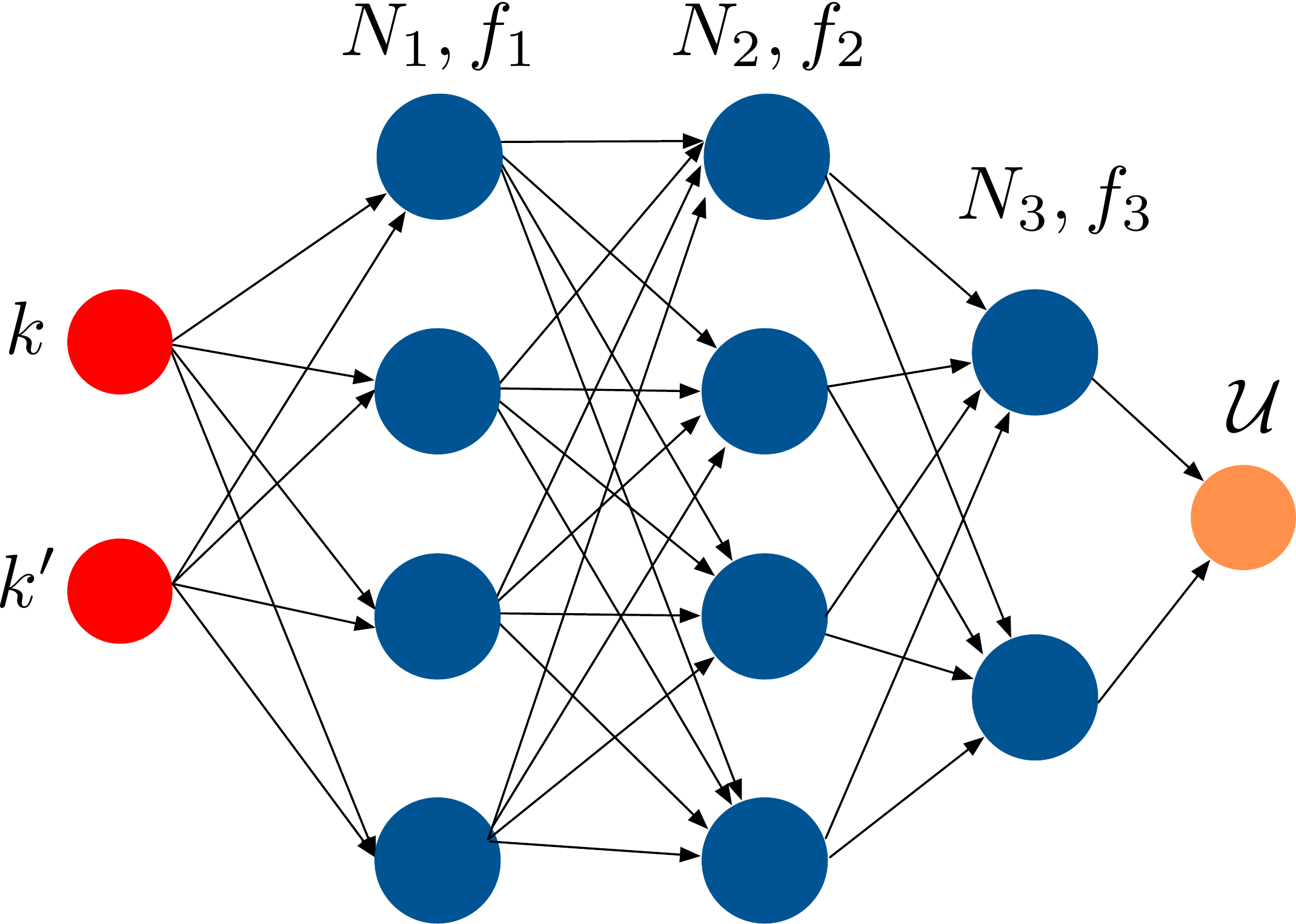}
\end{center}
\caption{Schematic of the fully connected neural network used in these two examples.}
\label{network}
\end{figure}

\section{Three-body Problem} 
In the first example, we consider a quantum three-boson problem. These bosons interact with a two-body pairwise potential described by an $s$-wave scattering length $a_\text{s}$ and a high-energy cut-off $\Lambda$. When $a_\text{s}$ is positive, there exists a two-body bound dimer state, and it is important to compute the atom-dimer scattering length $a_\text{ad}$ by solving the three-body problem. Here a key quantity is the scattering kernel $\mathcal{U}({\bf k}, {\bf k^\prime})$. Focusing on the $s$-wave scattering only, $\mathcal{U}$ only depends on the amplitude of momentum $k=|{\bf k}|$ and $k^\prime=|{\bf k}^\prime|$, and we can simplify $\mathcal{U}$ as $\mathcal{U}(k,k^\prime)$. Once we know $\mathcal{U}(k,k^\prime)$, one can calculate $a_\text{ad}$ through the Skorniakov-Ter-Martirosion (STM) equation \cite{Braaten}. The details of how to compute $\mathcal{U}(k,k^\prime)$ and how to obtain $a_\text{ad}$ from $\mathcal{U}(k,k^\prime)$ are summarized in Appendix A. 

\begin{figure}[t]
\begin{center}
\includegraphics[width=0.45\textwidth]{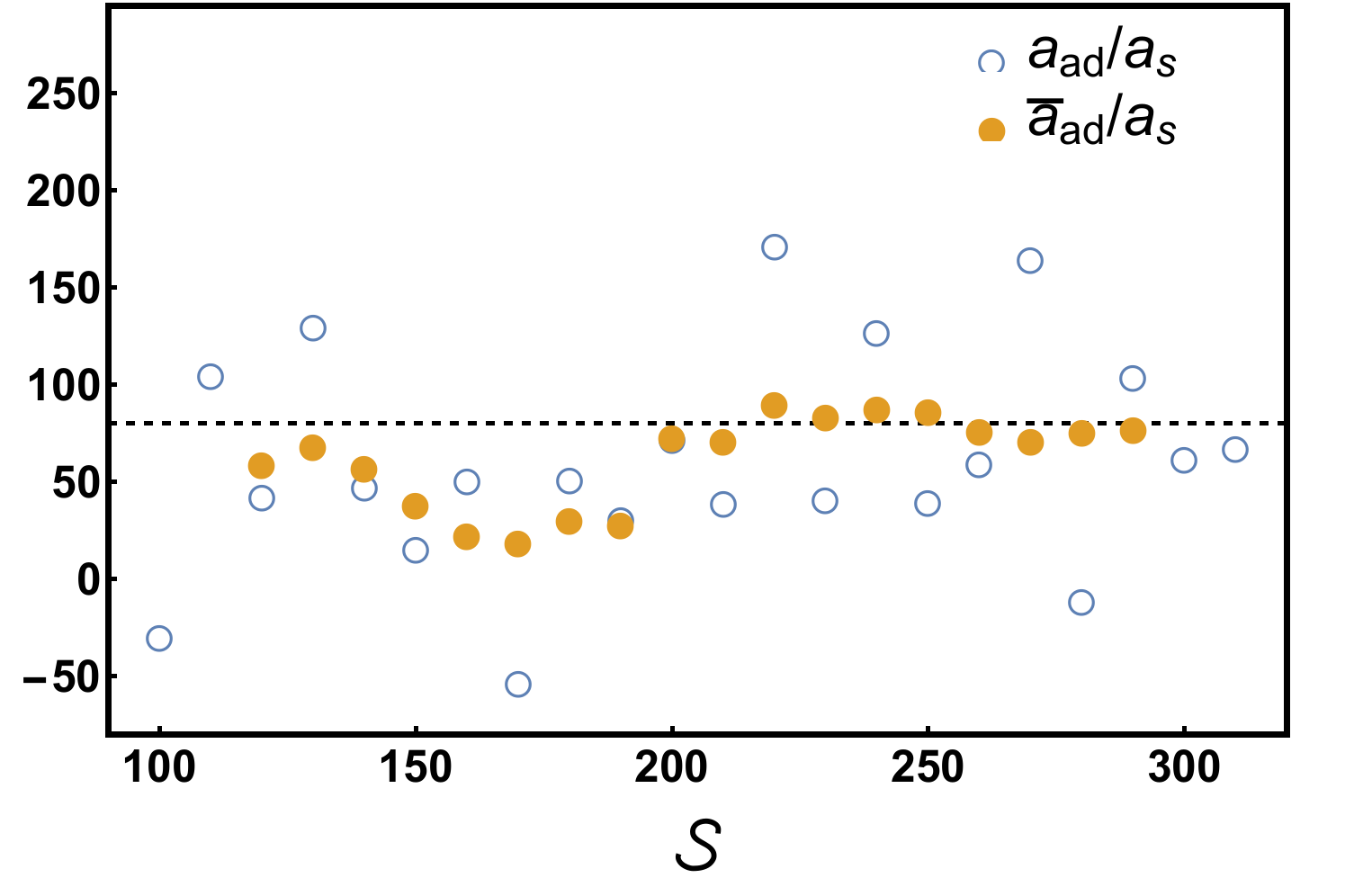}
\end{center}
\caption{The atom-dimer scattering length $a_\text{ad}$ calculated with the active learning method. $a_\text{ad}$ converges with the increasing number of queried dataset $\mathcal{S}$. The blue empty circles are the results averaged over different NNs at each step, and the yellow solid dots are results averaged over the adjacent five steps. The dashed line denotes the exact results obtained with all $10^4$ data points. Here we take the number of initial dataset $\mathcal{S}_0=100$ and at each step $\mathcal{S}_t=10$ data points are added.  $a_\text{s}\Lambda=10$. }
\label{performance}
\end{figure}

In the conventional method, we uniformly discretize both $k$ and $k^\prime$ into $L=100$ points between zero and $\Lambda$, and we need to compute $\mathcal{U}$ for all $10^4$ data points.
We then solve the STM equation with these $U(k,k^\prime)$ to obtain $a^{\rm exact}_\text{ad}$, and we have checked that such a discretization can ensure reaching the convergence, which is referred as the exact result and is shown by the dash line as reference in the Fig. \ref{performance}. Here we will follow the general procedure described above to fit $\mathcal{U}(k,k^\prime)$ using NNs. We will show that i) at most only $300$ number of data points are needed in order to obtain a reliable fitting, and ii) most of the queried data occupy the parameter regime where the function $\mathcal{U}(k,k^\prime)$ is steep, and iii) we use the trained NN to generate $\mathcal{U}(k,k^\prime)$ on all  $10^4$ data points to solve STM equation, and we find that the error is only few percent compared with the exact result.  

To be more concrete, we start with an initial dataset with uniformly sampled $S_0=100$ points. Here we design five different fully connected NNs, whose structures are schematically shown in Fig. \ref{network}. The input of all NN are two numbers $k$ and $k^\prime$, and the output is $\mathcal{U}$. Each layer of a NN is characterized by the number of nodes $N_\alpha$ and an activation function $f_\alpha$, and we describe each NN with by $(N_1, f_1; N_2,f_2; \dots)$. The five different NN used in this work are 
\begin{equation}\begin{aligned}
1:& ~(20, \tanh;20,\tanh;6,{\rm LS}) \\
2:& ~(20, \tanh;20,{\rm LS};6,\tanh) \\
3:& ~(30, \tanh;20,\tanh;6,{\rm LS}) \\
4:& ~(30, \tanh;20,{\rm LS};4,\tanh) \\
5:& ~(30, \tanh;20,\tanh;10,{\rm LS};4,\tanh),
\label{EqFNNs}
\end{aligned}\end{equation} 
where ${\rm LS}$ denotes the logistic sigmoid activation function with $f(a)=1/(1+e^{-a})$. It is important to keep these NNs different but their detailed structures are not important for final results. For each NN, the training results also depend on the initialization, which is particularly so when the number of data points are not enough. Therefore, for each NN, we also consider $20$ different initialization.

\begin{figure}[t]
\begin{center}
\includegraphics[width=0.47\textwidth]{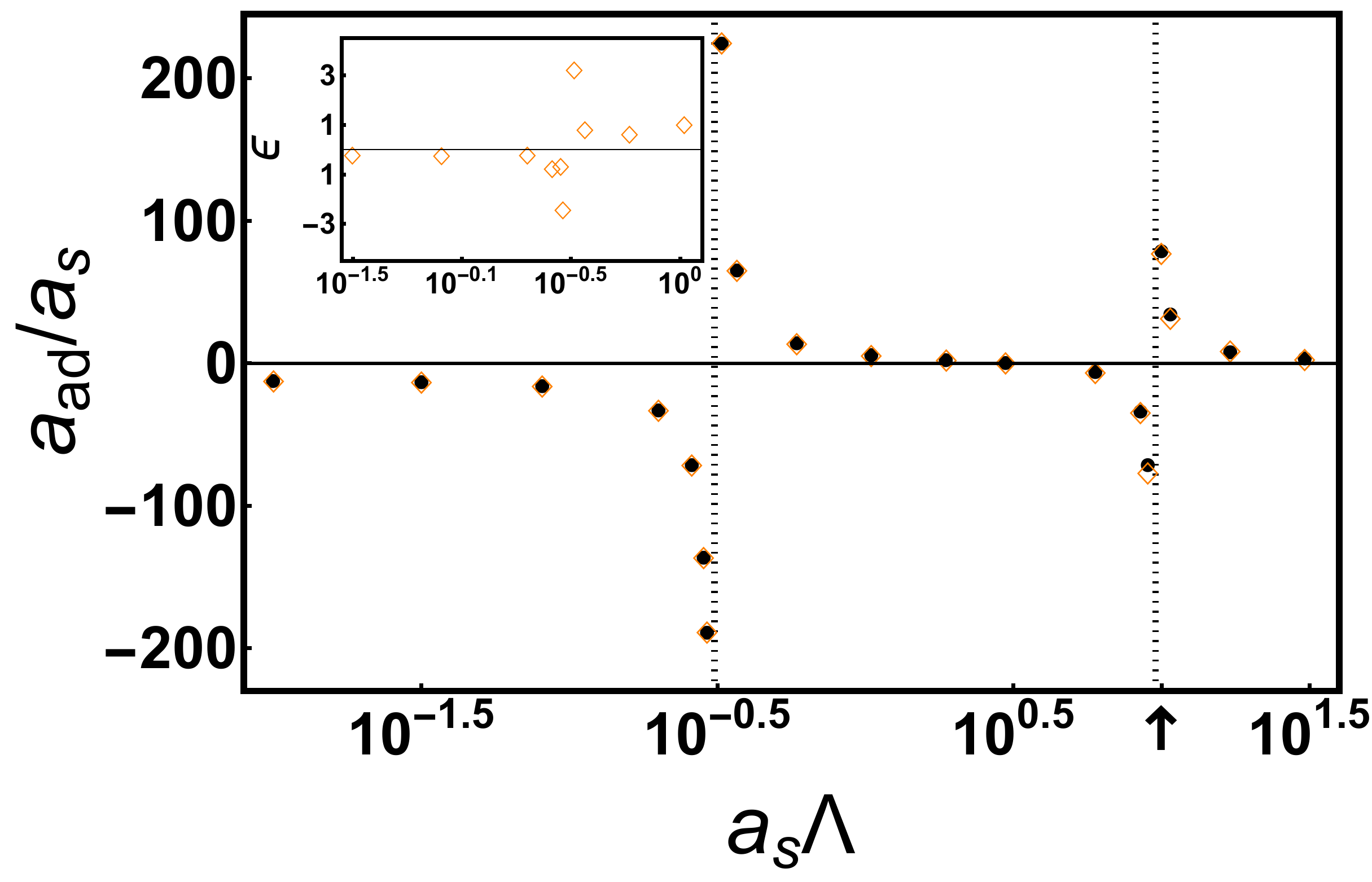}
\end{center}
\caption{$a_\text{ad}$ as a function of $a_\text{s}\Lambda$. The divergence of $a_\text{ad}$ is the atom-dimer resonance due to the Efimov effect. The black dots are calculated by evaluating $\mathcal{U}$ with uniform sampling all data points, and the yellow diamonds are results from the active learning algorithm with self-averaging. Arrow mark the $a_\text{s}\Lambda$ where the training processes are demonstrated Fig. \ref{performance}. The inset plots the relative error $\epsilon$\textperthousand ~ 
around the first atom-dimer resonant point.}
\label{Efimov}
\end{figure}

Therefore, at each iteration, we totally have $5\times 20$ trained NN to predict $\mathcal{U}$ on all $\{k,k^\prime\}$ points in the set $\mathcal{M}$. For each point $i \equiv \{k,k^\prime\}$, we can compute variance $\sigma_i$ for all $100$ predications. We will select $\mathcal{S}_\text{t}=10$ points with the largest variance to compute $\mathcal{U}$ at these points, and add them into the training set.  At each iteration, we can compute the mean variance $\sigma_{\rm mean}=1/L^2\sum_{i\subset \mathcal{M}}\sigma_{i}$. When the mean variance $\sigma_{\rm mean}$ is small enough, we start to evaluate the physical quantity $a_{\rm ad}$.

Here it comes to another important consideration of our method. On one hand, we have utilized the discrepancy between NNs to guide the query processes more efficiently, but on the other hand, when we start to compute observables, we need to properly average out these variances between different NNs to obtain a converged result. Hence, we compute $a_\text{ad}$ as follows:

1) For each NN, since there are $20$ copies depending on different initializations, we first average the predications over these $20$ copies to obtain a mean value $\bar{\mathcal{U}}_{\rm t}({k,k^\prime})$ for each point. 
We then choose one of $\mathcal{U}_{\rm t }^{j}({k,k^\prime})$ that is the closest to $\bar{\mathcal{U}}_{\rm t}({k,k^\prime})$, 
and we use this $\mathcal{U}_{\rm t }({k,k^\prime})$ (ignoring the upper index) as the representative of the $t$-th NN. 

2) We solve the STM equation with $\mathcal{U}_{\rm t }({k,k^\prime})$ and obtain $a^t_\text{ad}$. Among all five $a^t_\text{ad}$, we discard the largest and the smallest one, and take an average over the rest three, which is taken as $a_\text{ad}$ predicted by the active learning approach. The results are shown in Fig. \ref{performance} with blue circles. One can see that the results fluctuate around the exact value, and the fluctuation decreases as $\mathcal{S}$ increases. 

3) To further suppress the fluctuation, we can further take a self-average of $a_\text{ad}$, that is to average over $a_\text{ad}$ for five successive iterations. This result is denoted by $\bar{a}_\text{ad}$ and shown in Fig. \ref{performance} with yellow solid dots. Indeed, one can see that the fluctuation is already suppressed strongly at $\mathcal{S}\sim 200$.

When the result does not change with training epoch, we take a self-average over the last five iterations to obtain $\bar{a}_\text{ad}$, and we compare this result with $a^{\rm exact}_{\rm ad}$. For instance, when $a_s\Lambda=10$, the relative error $(\bar{a}_{\rm ad}-a^{\rm exact}_{\rm ad})/a^{\rm exact}_{\rm ad}$ is about $1\%$.

We apply this method to scan different values of $a_\text{s}$. We stop the iteration at $\mathcal{S}=300$, a number much smaller than uniformly sampled $10^4$ data points, and we take a self-average over the last five iterations to obtain $\bar{a}_\text{ad}$. The results are shown in Fig. \ref{Efimov}. Our calculation can obtain the right location for the atom-dimer resonances, and our active learning method can well reproduce the Efimov scaling law. By compared with the exact result, we define a relative error as $(\bar{a}_{\rm ad}-a^{\rm exact}_{\rm ad})/a^{\rm exact}_{\rm ad}\equiv\epsilon$\textperthousand. The $\epsilon$ around the first resonance is shown in the inset of Fig. \ref{Efimov}, which shows that the relative error does increases nearby the Efimov resonance, but overall it is kept within a few thousandths.

\begin{figure}[t]
\begin{center}
\includegraphics[width=0.45\textwidth]{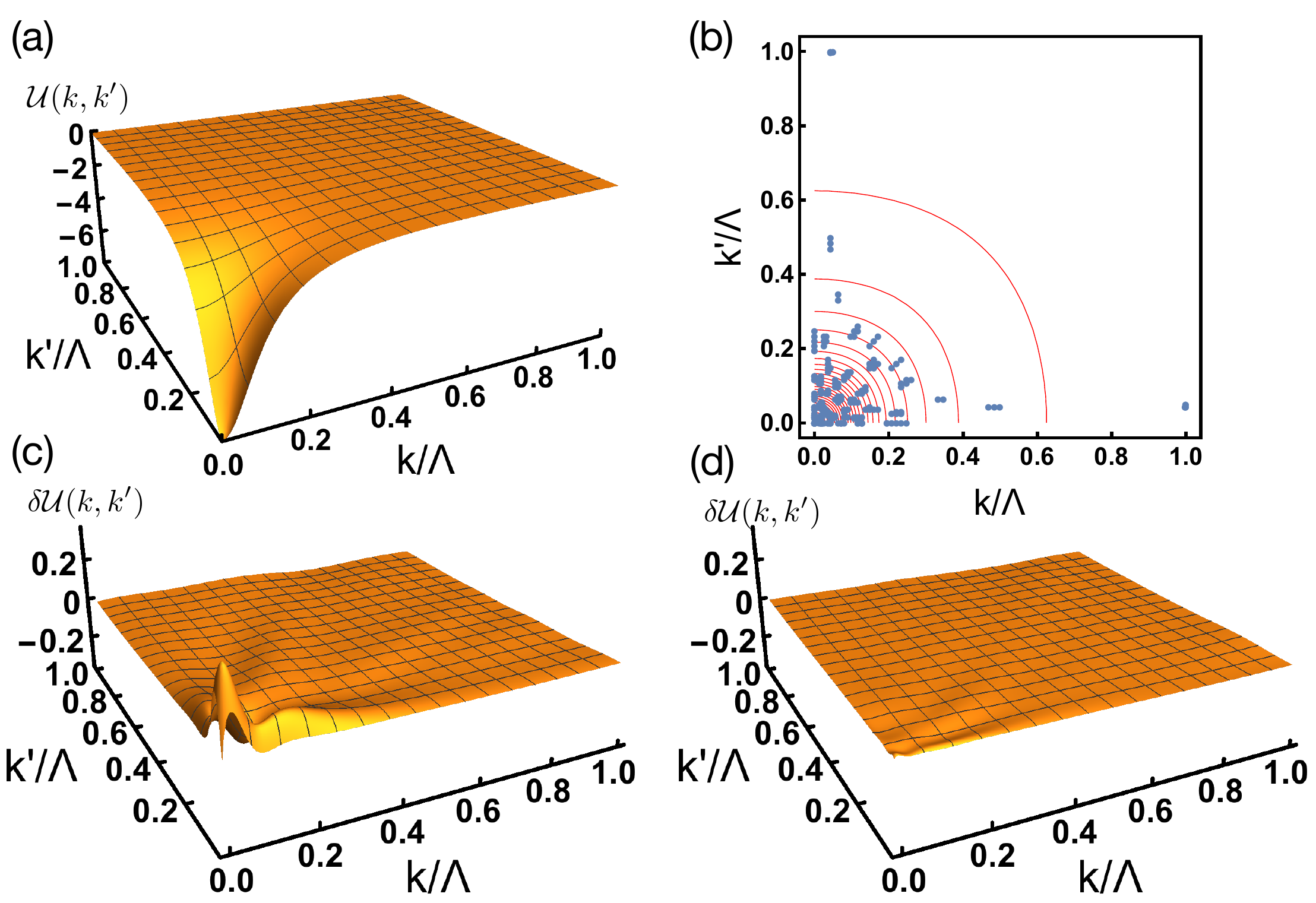}
\end{center}
\caption{ (a) Profile of the two-dimensional function $\mathcal{U}(k,k^\prime)$ in the whole momentum space, plotted with all uniformly sampled $10^4$ data points. (b) Visualization of the dataset queried during the active learning iterations.  The blue dots denote the queried dataset with $100< \mathcal{S} \leqslant 300$. The red lines are the contour-plotting of (b). (c-d) Difference between function $\mathcal{U}(k,k^\prime)$ generated by the NNs and $\mathcal{U}(k,k^\prime)$ generated by uniform sampling. Here we have averaged over all five different NNs and twenty different initializations. The labelled data $\mathcal{S}=120$ for (c) and $\mathcal{S}=320$ for (d).  }
\label{function_U}
\end{figure}

To further reveal the mechanism of how our method works, we plot in Fig. \ref{function_U}(a) the function $\mathcal{U}(k,k^\prime)$ generated by the uniform sampling of all $10^4$ points. All the data points added during iteration (with $100<\mathcal{S}\leqslant300$) are shown in Fig. \ref{performance}(b). It is very clear that nearly all points are added in the regime around $k\sim 0$ and $k^\prime \sim 0$, where $U(k,k^\prime)$ is the steepest as one can see from Fig. \ref{function_U}(a) and the equal number contour in Fig. \ref{function_U}(b). In Fig. \ref{function_U}(c) and (d), we compare the function $\mathcal{U}(k,k^\prime)$ generated by NNs and the function generated by uniform sampling. Here the NN results are averaged over five different NNs and different initializations of different NNs. It shows that the fitting is perfect when the calculation of $a_\text{ad}$ converges. 

\begin{figure}[t]
\begin{center}
\includegraphics[width=0.47\textwidth]{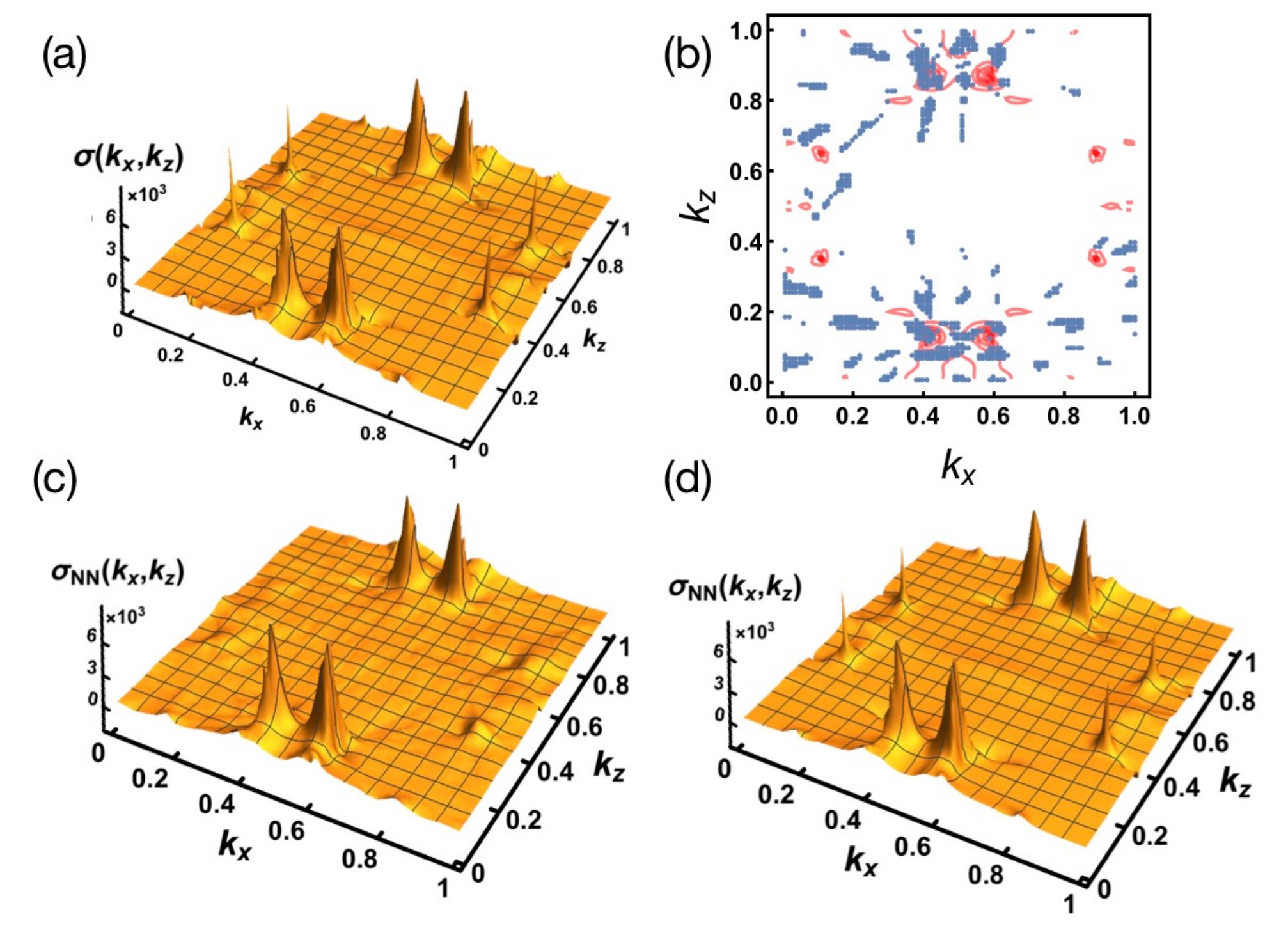}
\end{center}
\caption{
(a) Profile of the two-dimensional function $\sigma(k_x,k_z)$ in the whole momentum space, plotted with all uniformly sampled $10^4$ data points. (b) Visualization of the dataset queried during the active learning iterations.  The blue dots denote the queried dataset with $100< \mathcal{S} \leqslant 900$. The red lines are the contour-plotting of (b). $\sigma_\text{H}$ is in unit of $\text{S/cm}$. (c-d) The function $\sigma(k_x,k_z)$ generated by the NNs, averaged over all five different NNs and five different initializations for each NN. The labelled data $\mathcal{S}=1200$ for (c) and $\mathcal{S}=4000$ for (d). 
 }
\label{function_sigma}
\end{figure}

\section{Anomalous Hall Conductivity Problem}
In the second example, we consider the anomalous Hall conductivity of a magnetic Weyl semimetal Mn$_3$Ge \cite{Kubler2014}, which has been extensively studied recently \cite{Nakatsuji2015,Nayak2016}.
The anomalous Hall conductivity ($\sigma_{\text{Hall}}$) is an intrinsic quantity induced  by the Berry curvature of the band structure. We have computed $\sigma_{\text{H}}=\sum_{\mathbf{k}} \sigma(k_x,k_y,k_z)$  based on the $ab~initio$ band structure calculations. The model of this material and the details of computing the Berry curvature are shown in Appendix B.  
For each $k_x$ and $k_z$ point, we can obtain a value $\sigma(k_x,k_z)$ by performing an integration over $k_y$. Here without the active learning method, when we discretize both $k_x$ and $k_z$ into $100$ points, there are totally $10^4$ data points to be computed. The total Hall conductivity is obtained by summing over $k_x$ and $k_z$ as $\sigma_{\text{H}}=\sum_{k_x,k_z} \sigma(k_x, k_z)$. 

\begin{figure}[t]
\begin{center}
\includegraphics[width=0.47\textwidth]{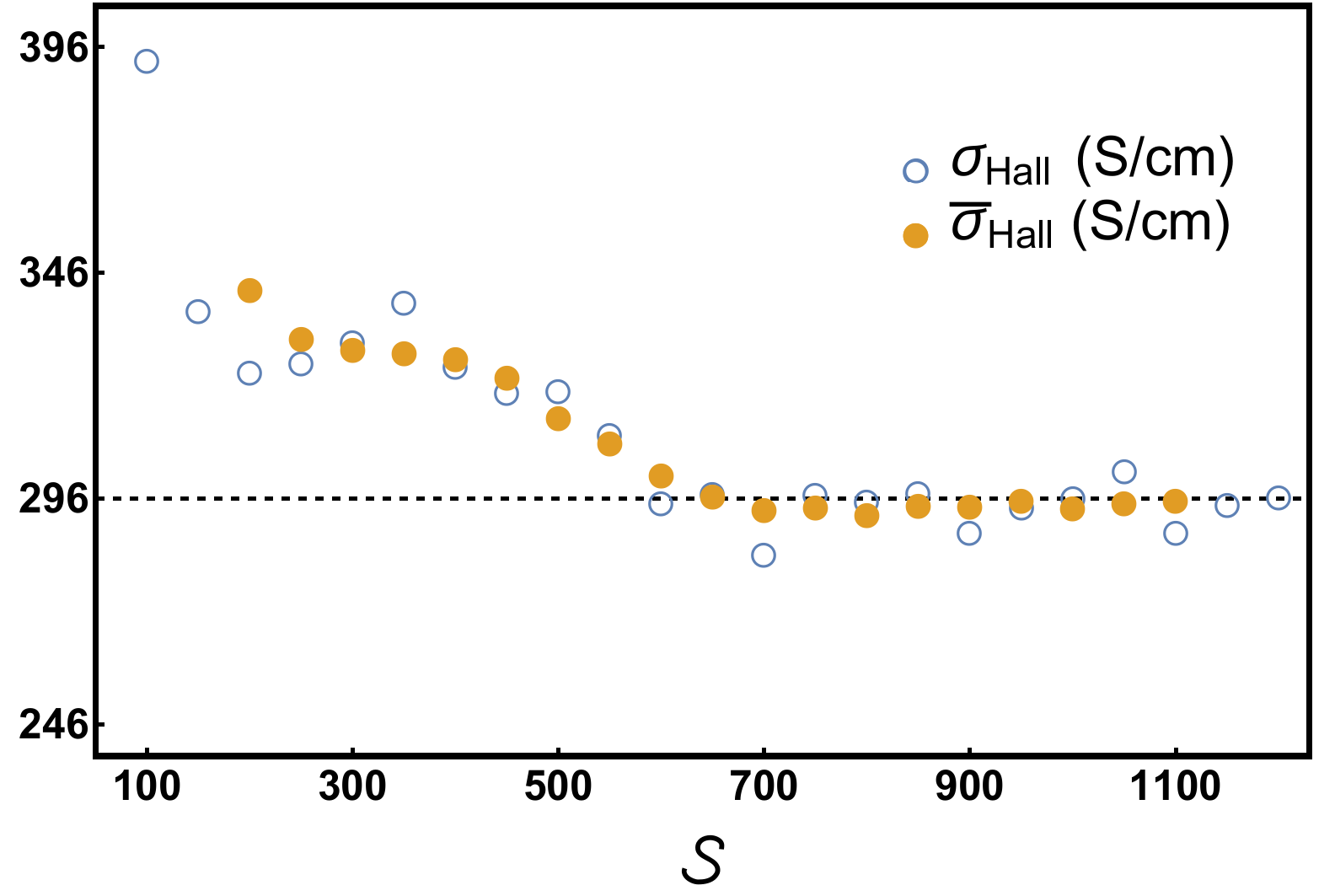}
\end{center}
\caption{The Hall conductivity $\sigma_\text{H}$ calculated with the active learning method. $\sigma_\text{H}$ converges with the increasing number of queried dataset $\mathcal{S}$. The blue empty circles are the results averaged over different NNs at each step, and the yellow solid dots are results averaged over the adjacent five steps. The dashed line denotes the exact results obtained with uniformly sampled all $10^4$ data points. Here we take the number of initial dataset $\mathcal{S}_0=100$ and at each step $\mathcal{S}_t=50$ data points are added. 
 }
\label{Hall}
\end{figure}

As one can see from Fig. \ref{function_sigma}(a), the function $\sigma(k_x,k_z)$ is much more singular than $\mathcal{U}(k,k^\prime)$ in the previous case. There are four broad peaks and four narrow peaks located around $(0.5\pm0.1,0.5\pm0.4)$ and $(0.5\pm0.4,0.5\pm0.15)$. We follow the same active learning procedure discussed above. In this case we take five different initializations for each NN. Compared with the first example, computing physical observable is easier because we only need to average over all five different initializations for each NN. Similarly, to suppress the fluctuation, at each iteration, we also average over three different predictions of $\sigma_\text{H}$ by discarding the largest and the smallest results. We also take a self-average over five successive iterations.  As we show in Fig. \ref{Hall}, the prediction of the active learning method approaches the exact value when $\mathcal{S}_{\rm max}>700$, which is less than $10\%$ of the total number of uniformly sampled data, and the relative error is about $0.8$\textperthousand. As we shown in Fig. \ref{function_sigma}(b), the ``query by committee" stratagem guides most of the queried date distributed in the areas of four broad peaks. In Fig. \ref{function_sigma}(c) and (d), we show the function $\sigma(k_x,k_z)$ generated by NNs. For $\mathcal{S}=1200$ shown in Fig. \ref{function_sigma}(c), the fitting has already captured the four main peaks, and since the contribution to the Hall conductance mainly comes from the four main peaks, the results of $\sigma_\text{H}$ already converges very well. Nevertheless, when we continue to add labelled data until $\mathcal{S}=4000$, as shown in Fig. \ref{function_sigma}(d), one can see that these extra data are mainly added in the four narrow peaks such that these four small peaks can also be generated very well. 

\section{Conclusion and Outlook} 
In summary, we have developed a neural network based machine learning method to determine a multi-dimensional function efficiently. The method combines the great expressivity of NN and the advantage of the active learning scheme to reduce the demand for labeled data to a minimum. There are two key ingredients of this method. On the one hand, we make use of the variances between NNs to guide the queried data to be sampled into the regime where the function varies rapidly, which makes the calculation more efficiently. On the other hand, we need to properly average out these variances when calculating physical observables using the fitted results. With two examples, we show our method can work remarkably well. Compared with uniform sampling, our method can achieve an accuracy of less than $1\%$ error with only less than $10\%$ of total data points, even when the function has multiple sharp peaks. We believe our method can find broad applications in many areas of computational physics. 

\section{Acknowledge}
We thank colleagues in Microsoft Research Asia for discussing active learning. This work is supported Beijing Outstanding Young Scientist Program (HZ), MOST under Grant No. 2016YFA0301600 (HZ) and NSFC Grant No. 11734010 (HZ), the Willner Family Leadership Institute for the Weizmann Institute of Science (BY), the Benoziyo Endowment Fund for the Advancement of Science (BY), and Ruth and Herman Albert Scholars Program for New Scientists (BY).

\begin{appendix}

\section{Three-boson problem}

\begin{figure}[t]
\begin{center}
\includegraphics[width=0.45\textwidth]{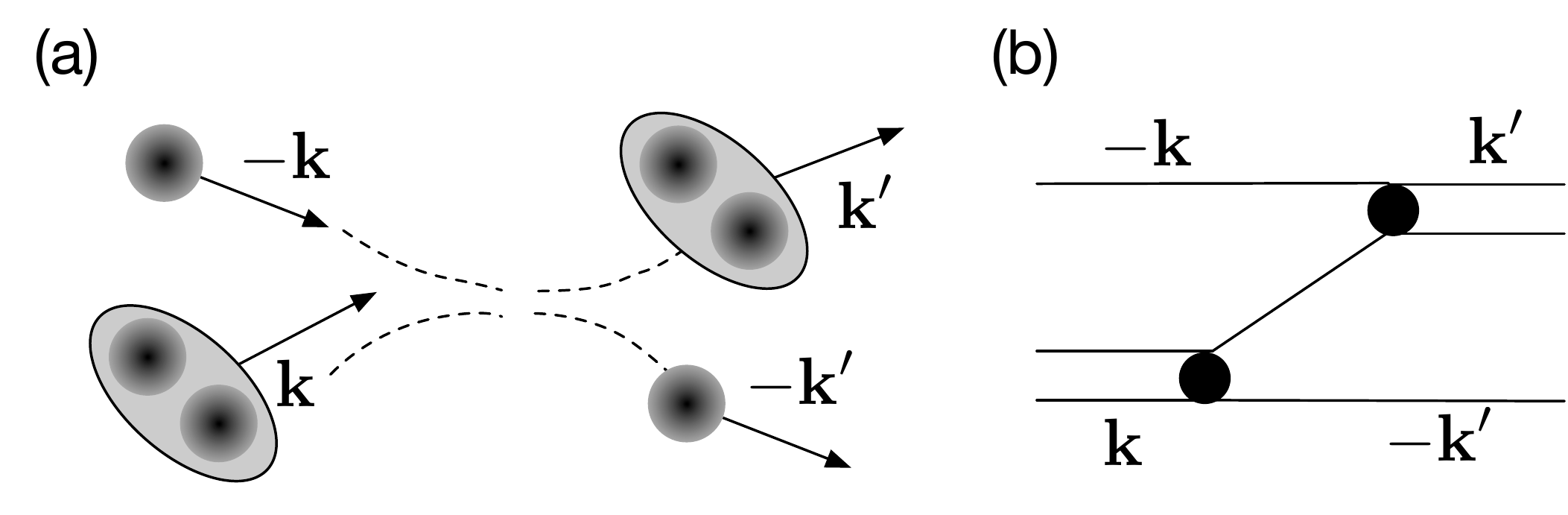}
\end{center}
\caption{(a) Schematic of the atom-dimer scattering process. (b) Diagrammatical description of the scattering kernel $\mathcal{U}({\bf k},{\bf k}^\prime)$. The single line stands for the free single-particle Green's function $G_0^{\rm A}({\bf k},E)$. The double line is the free dimer Green's function $D_s({\bf k},E)$. The black dot is the $s$-wave interacting vertex. }
\label{atom_dimer}
\end{figure}

The Hamiltonian of the three-boson system is given by
\begin{equation}
\hat{H}=\sum\limits_{i=1}^{3}-\frac{\nabla^2_i}{2m}+\sum\limits_{i<j,=1}^3V(|{\bf r}_i-{\bf r}_j|),
\end{equation}
where ${\bf r}_i$ ($i=1,2,3$) is the spatial coordinate of the $i$-th particle, and $V(r)$ is an isotropic short-range potential. Normally we focus on the $s$-wave scattering for ultracold low-energy atoms, and $V(r)$ can be described by the $s$-wave scattering length $a_\text{s}$ and a  high-momentum  cut-off $\Lambda$. Most features of this quantum three-body problem only depend on the dimensionless parameter $a_\text{s}\Lambda$ \cite{Braaten}. We consider the case that $a_\text{s}$ is large and positive, where $V(r)$ supports a low-energy bound state called a dimer. A key process to this three-body problem is the atom-dimer scattering illustrated in Fig. \ref{atom_dimer}(a), and the quantity to describe this process is a scattering kernel, which is usually denoted by $\mathcal{U}({\bf k},{\bf k}^\prime)$, and it can be computed  diagrammatically with the Feynman diagram shown in Fig. \ref{atom_dimer}(b) \cite{Braaten}. When focusing on the $s$-wave scattering only, $\mathcal{U}$ only depends on the amplitude of $k=|{\bf k}|$ and $k^\prime=|{\bf k}^\prime|$, and we can simplify $\mathcal{U}$ as $\mathcal{U}(k,k^\prime)$.

Knowing $\mathcal{U}(k,k^\prime)$, one can calculate the few-body quantity, such as atom-dimer scattering length $a_\text{ad}$, through the celebrated Skorniakov-Ter-Martirosion (STM) equation, which can be written as
\begin{align}
\int\frac{d k^\prime k^{\prime2}}{2\pi^2}\mathcal{U}(k,k^\prime)D_s\left(k^\prime,E\right)\mathcal{A}(k^\prime)-\mathcal{A}(k)=\mathcal{U}(k,0),
\label{EqSTM}
\end{align}
where 
\begin{align}
D_s^{-1}\left(k^\prime,E\right)=\frac{m}{2\pi a_\text{s}}-\frac{m}{2\pi}\sqrt{\frac{3k^{\prime 2}}{4}-mE},
\label{EqU}
\end{align}
is the full propagator for $s$-wave dimer. 
Solving Eq. \ref{EqSTM} with $E$ fixed at the dimer energy $-\frac{1}{ma_\text{s}^2}$, we can obtain atom-dimer scattering amplitude $\mathcal{A}(k)$, and $a_\text{ad}$ is given by \cite{Braaten}
\begin{equation}
a_{\rm ad}=-\frac{8}{3ma_s}\mathcal{A}(k=0).
\end{equation}

In practices, suppose that we discretize both $k$ and $k^\prime$ into $L=100$ points between zero and $\Lambda$, Eq. \ref{EqSTM} becomes a matrix equation after the discretization, which can be solved by inverting the matrix. Let $\mathcal{M}$ denotes the set of total data points, and the number of data points in $\mathcal{M}$ is $10^4$. With uniform sampling, we need to compute $\mathcal{U}$ for all $10^4$ points, with which we solve Eq. \ref{EqSTM} to obtain $a^{\rm exact}_\text{ad}$. This is referred as the exact results in the main text.

\section{Anomalous Hall conductivity}
The band structure of Mn$_3$Ge calculated with the density-functional theory from Vienna \textit{ab-initio} simulation package \cite{kresse1996} in the framework of the generalized-gradient approximation. The calculated plane wave basis result were projected to atomic-orbital-like Wannier functions \cite{Mostofi2008} to get the tight-binding parameters ($t_{ij}$). Based on the tight-binding Hamiltonian, 
\begin{flalign}
\hat{H} = \sum_{i,j}t_{ij}c^{+}_i c_j,
\end{flalign}
we calculated the Berry curvature and the anomalous Hall conductivity in the clean limit. 
 For the sufficient data set, uniform K point grid as $100\times100\times100$ employed to produce $10^6$ numbers of data points. We have evaluated the AHC ($\sigma_{zx}$) and Berry curvature ($\Omega_{zx}$) by the Kubo-formula approach in the linear response scheme \cite{Xiao2010},

\begin{flalign}
    \sigma_{xz}(\mu)&= -\frac{e^2}{\hbar}\int_{BZ} \frac{d\textbf k}{(2\pi)^3}\sum_{\epsilon_n<\mu}\Omega^n_{xz
    }(\textbf k) \\ 
    \Omega_{xz}^n(k)&= i \sum_{m\ne n } \frac{\langle n|\hat{v}_x|m \rangle \langle n|\hat{v}_z|m \rangle - (x\leftrightarrow z) }{(\epsilon_n\textbf{(k)} - \epsilon_m\textbf{(k)})^2}.
\end{flalign}

Here $n, m$ are the band index, $\epsilon_n$ is the eigenvalue of the $|n\rangle$ eigenstate of $H_k$ (Fourier transform of $\hat{H}$), and $\hat{v}_i=d H_k/(\hbar dk_i)$ ($i=x,z$)  is the velocity operator, $\mu$ is Fermi level of the system. 
The calculated anomalous Hall conductivity value is 296 S/cm where S denoting unit Siemens.

\end{appendix}

\end{document}